\documentclass{article}

\usepackage[english]{babel}
\usepackage{amsmath,amssymb}

\title{CP Violation in charm decays at CDF}
\author{A.~Di~Canto\footnote{Speaker on behalf of the CDF Collaboration.}}
\date{Physikalisches Institut, Ruprecht-Karls-Universit\"{a}t Heidelberg, Germany, and INFN Pisa, Italy.}

\newcommand{\A}{\ensuremath{\mathcal{A}}}
\newcommand{\Acp}{\ensuremath{\A_\textit{CP}}}
\newcommand{\acp}[1]{\ensuremath{\Acp^{\text{#1}}}}
\newcommand{\stat}{\ensuremath{\mathrm{~(stat)}}}
\newcommand{\syst}{\ensuremath{\mathrm{~(syst)}}}
\newcommand{\Dbar}{\ensuremath{\overline{D}{}}}

\begin{document}

\maketitle

\begin{abstract}
Exploiting the full Run II data sample collected by the CDF trigger on displaced vertices, we present a search for CP violation in neutral $D$ mesons decays to hadronic final states. We use the strong $D^{*+}\to D^0\pi^{+}$ (and c.c.) decay to identify the flavor of the charmed meson at production time and exploit CP-conserving strong $c\bar{c}$ pair-production in $p\bar{p}$ collisions. The results are the world's most precise measurements to date and confirm the presence of sizable CP-violating effects in the charm sector as recently observed by the LHCb collaboration.
\end{abstract}

\vskip30pt

While CP violation is well established for $B$ and $K$ mesons, this is not yet the case for charm mesons. First evidence for CP violation in two-body singly-Cabibbo-suppressed $D^0$ decays has been recently reported by the LHCb Collaboration \cite{lhcb}. Whether this is a hint of possible new physics contributions to the decay amplitude or not is not yet clear. It is important to broaden our search for CP violation in further charmed meson decays.

Owning to the large production cross-section available at the Tevatron collider and to the flexibility of a trigger for fully hadronic final states \cite{svt}, the CDF experiment, in a decade of successful Run II operations, collected millions of $D$ mesons decays, which allow high-precision CP-violation searches. Here we present two new measurements of CP violation in neutral $D$ mesons decays which are among the world's most sensitive to date. In all cases, the production flavor of the neutral $D$ meson is tagged by the charge of the pion from the $D^{*+}\to D^0\pi^+$ decay (charge conjugated states are implied, unless otherwise stated).

\section{Time-integrated asymmetries in $D^0\to K^0_S\pi^+\pi^-$ decays}
In a data sample corresponding to an integrated luminosity of $6$\,fb$^{-1}$, CDF searches for time-integrated CP asymmetries in the resonant substructure of the three-body $D^0\to K^0_S\pi^+\pi^-$ decay. As the Standard Model expectation of these CP asymmetries is $\mathcal{O}(10^{-6})$ \cite{bigi-xing}, well below the experimental sensitivity, an observation of CP violation would be a clear hint of new physics.

We reconstruct approximately $350\,000$ $D^*$-tagged $D^0\to K^0_S(\to\pi^+\pi^-)\pi^+\pi^-$ candidates. Two complementary approaches are used: a full Dalitz fit and a model-independent bin-by-bin comparison of the $D^0$ and $\Dbar^0$ Dalitz plots. We briefly present here only the result of the first approach, a more comprehensive description of the analysis can be found in Ref.~\cite{kspipi}.

\begin{table}[t]
\centering
\caption{Measured fit fraction asymmetries, $\A_{FF}$, for the considered intermediate resonances of the $D^0\to K_S^0\pi^+\pi^-$ decay. The first uncertainties are statistical and the second systematic.}\label{aff}
\begin{tabular}{lclc}
\hline
Resonance & $\mathcal{A}_{\mathrm{FF}}$ [\%] &  Resonance & $\mathcal{A}_{\mathrm{FF}}$ [\%] \\
\hline
$K^*(892)^-$ & $0.36 \pm 0.33 \pm 0.40$ & $K^*(892)^+$ & $1.0 \pm 5.7 \pm 2.1$ \\
$K_0^*(1430)^-$ & $4.0 \pm 2.4 \pm 3.8$ & $K_0^*(1430)^+$ & $12 \pm 11 \pm 10$ \\
$K_2^*(1430)^-$ & $2.9 \pm 4.0 \pm 4.1$ & $K_2^*(1430)^+$ & $-10 \pm 14 \pm 29$ \\
$K^*(1410)^-$ & $-2.3 \pm 5.7 \pm 6.4$ & $K^*(1680)^-$ & (not found)\\
$\rho(770)$ & $-0.05 \pm 0.50 \pm 0.08$ & $\rho(1450)$ & $-4.1 \pm 5.2 \pm 8.1$ \\
$\omega(782)$ & $-12.6 \pm 6.0 \pm 2.6$ & $\sigma_2$ & $-6.8 \pm 7.6 \pm 3.8$ \\
$f_0(980)$ & $-0.4 \pm 2.2 \pm 1.6$ & $f_0(1370)$ & $-0.5 \pm 4.6 \pm 7.7$ \\
$f_2(1270)$ & $-4.0 \pm 3.4 \pm 3.0$ & $f_0(600)$ & $-2.7 \pm 2.7 \pm 3.6$ \\
\hline
\end{tabular}
\end{table}

For the first time at a hadron collider, a Dalitz amplitude analysis is applied for the description of the dynamics of the decay. We employ the isobar model and determine the asymmetries between the different $D^0$ and $\Dbar^0$ sub-resonance fit fractions (Tab.~\ref{aff}) in order to be insensitive to any global instrumental asymmetry in the reconstruction and identification of the candidates of interest. The results represent a significant improvement in terms of precision with respect to previous determinations of these quantites \cite{cleo}, and no hints of any CP violating effects are found. The measured value for the overall integrated CP asymmetry is
$$\Acp(D^0\to K_S^0\pi^+\pi^-) = \bigl(-0.05\pm 0.57\stat\pm0.54\syst\bigr)\%.$$
Following the procedure described in Ref.~\cite{paper} and assuming no direct CP violation in the $D^0\to K_S^0\pi^+\pi^-$ decay ($\acp{dir}=0$), we can derive a measurement of time-integrated CP violation in $D^0$ mixing ($\acp{ind}$) since the measured time-integrated asymmetry can be approximately expressed as
\begin{equation}\label{acp}
\Acp(D^0\to f) \approx \acp{dir}(D^0\to f) +\frac{\langle t\rangle}{\tau}\ \acp{ind},
\end{equation}
where $f$ indicates a generic final state and $\langle t\rangle/\tau\approx2.28$ is the observed average $D^0$ decay time of the sample in units of $D^0$ lifetimes. We then find
$$\acp{ind} = \bigl(-0.02\pm 0.25\stat\pm0.24\syst\bigr)\%.$$

\section{Time-integrated asymmetries in $D^0\to K^+K^-$ and $D^0\to\pi^+\pi^-$ decays}
Building upon the techniques developed for the previous measurement of individual asymmetries in $D^0\to h^+h^-$ ($h=\pi$ or $K$) decays \cite{paper}, CDF updated and optimized the analysis toward the measurement of the difference of asymmetries, $\Delta\Acp = \Acp(D^0\to K^+K^-) - \Acp(\pi^+\pi^-)$. The difference of asymmetries is here used as a tool to cancel, to an excellent level of accuracy, the few percents detector-induced asymmetry in the efficiencies for reconstructing the tagging-pion from the $D^*$ decay. The offline selection has been loosened with respect to the measurement of individual asymmetries, since their difference is much less sensitive to instrumental effects allowing for a more inclusive selection, and we now use the full CDF Run~II data sample, which corresponds to $9.7$\,fb$^{-1}$ of integrated luminosity. Requirements on the minimum number of hits for reconstructing tracks are loosened, the $p_T$ threshold for $D$ decay products is lowered from $2.2$ to $2.0$\,GeV/$c$ and $\sim12\%$ fraction of charmed mesons produced in $B$ decays, whose presence does not bias the difference of asymmetries, is now used in the analysis. As a result of the improved selection, the $D^0$ yield nearly doubles and the expected resolution on $\Delta\Acp$ becomes competitive with LHCb's \cite{lhcb}. In the following we briefly present the result, more details can be found in Ref.~\cite{cdf10784}.

Using the approximately $550\,000$ $D^*$-tagged $D^0\to\pi^+\pi^-$ and $1.21\cdot10^6$ $D^*$-tagged $D^0\to K^+K^-$ decays, we measure
$$\Delta\Acp = \bigl(-0.62\pm0.21\stat\pm0.10\syst\bigr)\%,$$
which is $2.7\sigma$ different from zero and consistent with the LHCb result \cite{lhcb}, suggesting that CDF data support CP violation in charm. By means of Eq.~\eqref{acp} and using the observed values of $\langle t(K^+K^-)\rangle-\langle t(\pi^+\pi^-)\rangle = (0.27\pm0.01)\ \tau$, the observed asymmetry is combined with all other available measurements of CP violation in $D^0\to h^+h^-$ decays to extract the values of $\acp{ind}$ and $\Delta\acp{dir}=\acp{dir}(D^0\to K^+K^-)-\acp{dir}(D^0\to \pi^+\pi^-)$. The combination yields $\Delta\acp{dir} = (-0.67\pm0.16)\%$ and $\acp{ind} = (-0.02\pm0.22)\%$, which deviates by approximately $3.8\sigma$ from the no-CP violation point.

Finally, the measured value of $\Delta\Acp$ from the subsample of additional events selected by the new criteria is combined with the statistically independent results of Ref.~\cite{paper}, to obtain a more precise determination of the individual asymmetries:
\begin{align*}
\Acp(D^0\to\pi^+\pi^-) &= \bigl(+0.31\pm0.22\ (\text{stat.}+\text{syst.})\bigr)\%,\\
\Acp(D^0\to K^+K^-) &= \bigl(-0.32\pm0.21\ (\text{stat.}+\text{syst.})\bigr)\%.
\end{align*}

These results are the world's most precise to date; they improve and supersede the previous corresponding results of Ref.~\cite{paper}.

\end{document}